\documentclass[conference]{IEEEtran}
\IEEEoverridecommandlockouts

\usepackage[pdftex]{graphicx}
\usepackage{cite}
\usepackage{amsmath,amssymb,amsfonts}
\usepackage{algorithmic}
\usepackage{graphicx}
\usepackage{textcomp}
\usepackage{epsfig, algorithm}
\usepackage{epstopdf}
\usepackage{xcolor}
\usepackage{subfigure} 
\usepackage{graphicx}
\usepackage{epstopdf}
\usepackage{graphics}
\usepackage{epsfig, algorithm}
\usepackage{titlesec}
\usepackage{caption}
\usepackage{multirow}
\usepackage{cite}
\usepackage{array}
\usepackage{multirow}
\usepackage{tabularx}
\usepackage{multirow}
\usepackage[normalem]{ulem}
\usepackage[utf8]{inputenc}
\useunder{\uline}{\ul}{}
\usepackage[utf8]{inputenc}
\usepackage{placeins}
\usepackage[caption=false]{subfig}
\usepackage{balance}
\begin{document}

\title{Energy-Efficient Resource Allocation for 5G Cognitive Radio NOMA Using Game Theory\\}

\author{
\IEEEauthorblockN{Shaima' S. Abidrabbu\IEEEauthorrefmark{1} and H\"{u}seyin Arslan\IEEEauthorrefmark{1}\IEEEauthorrefmark{2}}\\
\IEEEauthorblockA{\IEEEauthorrefmark{1}Department of Electrical and Electronics Engineering, Istanbul Medipol University, Istanbul, 34810 Turkey\\
}
\IEEEauthorblockA{\IEEEauthorrefmark{2}Department of Electrical Engineering, University of South Florida, Tampa, FL 33620 USA\\
Email: sabidrabbu@st.medipol.edu.tr, huseyinarslan@medipol.edu.tr}

}
\maketitle

\begin{abstract}

Cognitive radio non-orthogonal multiple access (CR-NOMA)  networks promise improved spectrum utilization and capacity in 5G networks. In this work, we aim to investigate efficient power allocation for the secondary users (SUs) in underlay CR-NOMA networks using a game-theoretic approach. We present a novel power allocation to CR-NOMA network from a game-theoretic perspective. First, we specify the utility function of the primary users (PUs) and SUs, and formulate the game as a non-cooperative game. Then, the existence and uniqueness of the Nash equilibrium (NE) are investigated. Finally, the sum utilities of SUs is maximized by optimal power allocation at the NE point. Simulation results provided that the proposed scheme outperforms the conventional method, providing up to 37.5\% increase in sum utilities of the SUs. 
\end{abstract}


\section{Introduction}
Increasing user requirements such as capacity and data rates have been the driving force behind evolving communication technologies. Cognitive radio (CR) and non-orthogonal multiple access (NOMA) are two promising intended to improve spectral efficiency and consequently, system capacity \cite{duan2011investment}, \cite{fang2016energy} in 5G communication. CR-NOMA networks are seen as a specific case of power-domain NOMA applied in CR, wherein the requirements of the secondary users (SUs) and primary users (PUs) are strictly met, resulting in improved system performance \cite{8352630}.% providing a good system performance 
\par

However, there are many critical challenges due to the severe interference caused by NOMA in CR networks, which are related to resource allocation (RA) and interference management. Various efforts have been made to investigate and facilitate these challenges \cite{8352630}, \cite{lv2018cognitive}.  
% which are playing central role in designing the network. 
RA and optimization aim for efficiently utilizing the resources in terms of different objectives like spectral and energy efficiency\cite{liu2017interference}. For example, the authors in \cite{zhang2016energy} showed that the energy efficiency of underlay CR-NOMA can be higher than that of cognitive radio networks (CRNs) with orthogonal multiple access (OMA) using sequential convex approximation method. In \cite{zeng2016power}, the authors proposed a novel power allocation algorithm for CR-NOMA, where the characteristics of the NOMA-based system had totally exploited for designing the RA algorithm. RA in CR-NOMA studied also in the literature for simultaneous wireless information and power transfer (SWIPT) scheme. For example in \cite{song2019joint}, time sensing studied as a critical constraint in the optimization problem. In \cite{mao2019power}, the authors proposed a  non-linear energy harvesting (EH) model to minimize the overall system power consumption. In \cite{zhao2019energy}, the multi-objective optimization problem based on EH and the quality of service (QoS) of the users in CR-NOMA considered. 

One of the biggest issues in CR-NOMA networks is the design of an optimal RA scheme, considering different (and often competing) objectives such as spectral efficiency, energy efficiency, and interference management \cite{zhou2018state}. Although the multi-objective optimization can be exploited to achieve sub-optimal solutions, the complexity of the designed algorithms may be very high. To cope with this complexity, we propose using game theory (GT) to design efficient RA in the proposed network. To the best of our knowledge, this is the first work to study RA in CR-NOMA using GT approach. 

% To cope with this complexity of the design algorithms, we suggest using game theory. Considering this, our approach employs game theory for the RA among SUs of the CR-NOMA network. 
% {notes from the phone}. 
% But this complexity becomes much less when we are dealing with the game theory. Based on this reason, the RA of CR-NOMA is designed using game-theoretic (GT). To the best of our knowledge, this is the first work to study RA in CR-NOMA using non-cooperative GT approach. 

\par
In this paper, the power allocation of an underlying CR-NOMA network using the GT approach is investigated which introduces the  GT to the RA in candidate 5G networks . More specifically, in this work, the game is represented as competition between the $kth$ SU which is trying to predict the other players' strategies to maximize his payoff and at the last find the Nash equilibrium (NE) of the game at the SU base station. As a result, the resource management mechanism of the proposed CR-NOMA-based GT approach will achieve high net utilities for all the SUs while maximizing the energy and spectrum efficiency by achieving their satisfactions. Our contributions are summarized as follows:
\begin{itemize}
\item  A novel non-cooperative power control framework for the CR-NOMA network is presented, where each SU selfishly optimizes its power allocation over the allocated resources to maximize its utility function. Moreover, we prove the existence, and provide the conditions for the uniqueness of the NE. The performance of the proposed approach is compared with OMA-based conventional non-cooperative GT power control. 
\item Since the interference level to the PU is usually modeled as strict   constraints for resource allocation optimization in the literature of CRNs. In this work, the interference to the PU is taken into the consideration by introducing the PUs as a part of the game where the number of PUs and their utilities are conducting in the overall system model.  
\item   A novel algorithm is proposed to reach the NE-point as an iterative algorithm. NE-point is the optimal response of all users in the game such that no player gains more utility by unilaterally deviating or changing his strategy, under the assumption that other player(s) strategies remain unchanged.
\end{itemize}

This paper is organized as follows. Section II describes the analytical model SUs' and PUs' perspectives. Section III presents the formulation of the power control problem as a strategic non-cooperative game; Section IV provides the existence and uniqueness of the NE point, and a novel algorithm introduced to reach the NE. Simulation results are presented in Section V. Finally, Section VI concludes this work and highlights the future perspective of the proposed work.
\section{ANALYTICAL MODEL}
The uplink scenario for CR-NOMA networks is considered where the network consists of a set of PUs and SUs with different base stations (BSs). The sets of SUs and PUs are denoted as $x_{P}=\{1,2, \ldots, N\}$ and $x_{S}=\{1,2, \ldots, K\}$, respectively, where $N$ and $K$ are the total numbers of PUs and SUs, respectively. Hence, multiple SUs can be served by one or more PUs, and all the users are equipped with a single antenna. In CR-NOMA, the network design is based on the scheduling scheme that facilitates simultaneous access of SUs and PU using the entire system bandwidth to transmit data with the help of superposition coding $(\mathrm{SC})$ and successive interference cancellation (SIC) decoding techniques. User multiplexing is executed in the power domain, subject to the constraint PU power budget $\mathbb{Q}_{n}^{\max }$ and $\mathrm{SU}$ total power $\rho_{k}^{\max },$ respectively. In the uplink, SIC is at base stations, therefore users need not be aware of the modulation and coding schemes employed by the other users. Furthermore, base stations have enough processing power to perform SIC. NOMA can also allow users
to transmit in uplink in a grant-free manner which reduces latency significantly \cite{sedaghat2018user}.

Hence, $\mathbb{Q}_{n}^{\max }$ and $\rho_{k}^{\max }$ represent the maximum transmission power for PU and SU, respectively.
The channel gain between the secondary BS and $kth$ SU, and between the $nth$ PU-BS and SU-BS are represented as $h_{sk}$, and $h_{ns}$, respectively. In this work, NOMA is applied only for SUs, whereas for PU its optional to apply NOMA. So, SUs channels can be sorted in the SU-BS as $0<\left|h_{s1}\right|^{2}\leq\left|h_{s2}\right|^{2}\leq\cdots\leq\left|h_{sK}\right|^{2}$.

\subsection{Secondary User Perspective Analysis}
The power allocated to the $kth$ SU is
$\rho_{k}$ such that $\rho_{1}<\rho_{2}<\ldots<\rho_{K}$, and the data rate achievable of the $kth$ SU can be represented as
\begin{equation}
   \mathcal{R}_{k}^{s}=\mathcal{B} \log _{2}\left(1+\gamma_{k}^{(s)}\right),
\end{equation}
where $\mathcal{B}$ is the system bandwidth and $\gamma_{k}^{(s)}$ is the signal to interference plus noise ratio (SINR) for the $kth$ SU , that represented as
\begin{equation}
\gamma_{k}^{(s)}=\frac{h_{s k}^{2} \rho_{k}}{\sum_{j=1 \atop j \neq k}^{K} h_{s j}^{2} \rho_{j}+\sum_{m}^{N} h_{s m}^{2} \rho_{m}+\sigma^{2}}, 
\end{equation}
where $\rho_{k}$ is the transmission power of the $kth$ $\mathrm{SU}$ and $\sigma^{2}$ is the variance of the additive Gaussian white noise (AWGN). $u_{k}$ is a novel utility function for the $kth$ SUs that, shows the aims of $kth$ SU to maximize its data rate with minimum transmission power. Based on the above aforementioned, the SU who has a minimum power will have maximum utility. $u_{k}$  represented as
\begin{equation}
 u_{k}=\frac{\mathcal{R}_{k}^{s}}{\rho_{k}}. 
\end{equation}
\subsection{Primary User Perspective Analysis}
The target SINR of the $nth$ PU is defined as $\bar{\gamma}_{n}=\frac{h_{ns}^{2} \rho_{n}}{\mathbb{Q}_{N}+\sigma^{2}},$ where $\mathbb{Q}_{N}$ represents the maximum possible interference caused by the SUs and other PUs at the $nth$ $\mathrm{PU}$ that can be tolerated, $\bar{\gamma}_{n}$ represents the least acceptable transmission quality of the $nth$ PU (i.e., the least acceptable SINR of the $nth$ PU), and $\rho_{n}$ is the $nth$ PU transmission power. 

In order to benefit from dynamic spectrum sharing, the $\mathrm{PU}$ is rewarded for allowing the SUs to use its spectrum. However, the transmission quality of the PU must always be satisfied, therefore $\gamma_{p}^{(n)}-\bar{\gamma}_{n} \geq 0$. The $\gamma_{p}^{(n)}$ is the SINR of the $nth$ PU and, it implies that the QoS requirement for the $nth$ $\mathrm{PU}$ is satisfied. 
As defined in \cite{5090010}, $u_{n}$ is the utility function of the $nth$ PU given by 
\begin{multline}
  u_{n}=\mathbb{Q}_{N}-\mu_{1}\left(\mathbb{Q}_{N}-I_{n}-I_{k}\right)^{2} u\left[\mathbb{Q}_{N}-I_{n}-I_{k}\right]\\
- \mu_{2}\left[e^{\left(I_{n}+I_{k}-\mathbb{Q}_{N}\right)}\right] u\left[I_{n}+I_{k}-\mathbb{Q}_{N}\right]  ,
\end{multline}
where $u(.)$ is the unit step function, $\mu_{1}$ and $\mu_{2}$ are positive pricing coefficients and $ I_{k}=\sum_{j=1 \atop j \neq k}^{K} h_{s j}^{2} \rho_{j}+\sum_{m=1}^{N} h_{s m}^{2} \rho_{m}$ and  $I_{n}=\sum_{j=1}^{K} h_{p j}^{2} \rho_{j}+\sum_{m=1 \atop m \neq n}^{N} h_{p m}^{2} \rho_{m}$ represent the interference caused by other users in the network on the $kth$ SU and $nth$ PU, respectively. 
From the utility of the PU, we can conclude that when the instantaneous SINR of the $nth$ PU is less than the target SINR of the PU, the $nth$ $\mathrm{PU}$ is significantly penalized because it does not achieve its target transmission quality. As well, the $nth$ PU can be penalized if its  instantaneous SINR is grater than its target SINR since this can cause unnecessary interference to the other users. 

\section{Game Formulation}
In order to investigate the power allocation of  CR-NOMA networks, a  non-cooperative power control game theory is introduced in this section. The non-cooperative power control game for the system can be formulated in a strategic form as follows:
\begin{itemize}
\item Players: $x$ is the total number of the users in the system i.e., PUs and SUs, $x=\{1,2, \ldots, N, 1,2,3, \ldots, K\}$. 
\item Action space: $\mathcal{P}=\mathbb{Q}_{1} \times \mathbb{Q}_{2} \times \ldots \times \mathbb{Q}_{N} \times \mathcal{P}_{1} \times \mathcal{P}_{2} \times$
$\mathcal{P}_{3} \times \ldots \times \mathcal{P}_{K}.$ Here, $\mathbb{Q}_{n}=[0,\mathbb{Q}_{n}^{\max }]$ represents the action set of the $nth$ PU, $\rho_{k}=[0,\rho_{k}^{\max }]$ represents the action set of the $kth$ SU. Furthermore, the action vector of all users is $\mathbb{P}=$ $\left[\mathbb{Q}_{1},\mathbb{Q}_{2},\ldots,\mathbb{Q}_{N},\rho_{1},\rho_{2},\rho_{3},\ldots,\rho_{K}\right],\quad\rho_{k}\in\rho_{K}\quad$, and $\mathbb{Q}_{N}\in\mathcal{Q}_{n}$. The action vector excluding the $kth$ SU for $k=\{1,2,\ldots,K\}$ is denoted by $\rho_{-K}$ and the action vector excluding the $nth$ $\mathrm{PU}$ for $n=\{1,2,\ldots . . N\}$ is denoted by $\mathbb{Q}_{-N}$. 
\item Utility function: The utility function represents the motivations of players in the game. Here, $u_{k}\left(\rho_{K}, \rho_{-K}\right),\forall k=1,2,3,4,\ldots,K$ denotes the utility function of the $kth$ $\mathrm{SU}$ and the utility function of the $nth$ $\mathrm{PU}$ represents as 
$u_{n}\left(\mathbb{Q}_{N}, \mathbb{Q}_{-N}\right), \forall n=1,2,3,.....,N.\quad$ $\quad$ Additionally, each user utility depends on not only its strategy but also on the other users strategies in the game. Therefore, the $\mathbb{Q}_{-N}$ and $\rho_{-K}$ are crucial in the maximization or minimization of the $nth$ PU and $kth$ SU utility, respectively.  
\end{itemize}
\section{Existence and Uniqueness of the NE Point}
Generally speaking, NE is a concept of game theory where the optimal outcome of a game is one where no player has an incentive to deviate from his chosen strategy after considering an opponent's choice. NE is the best solution for the non-cooperative game power control game \cite{maskin1999nash}. In this section, the NE solution of the formulated game model is disclosed and investigated by the existence and uniqueness. 

According to the theorem of the existence of the NE shown in \cite{li2008novel}, the NE exists in a game $\mathcal{G}=(x,\mathcal{P},\mathcal{U})$, for all players $x=\{1,2,\ldots,N, 1,2,3,\ldots,K\},$ where $\mathcal{P}$ is a non-empty, convex, and compact subset of some Euclidean space $\mathbb{R}^{N+K}$, $\mathcal{U}$ is continuous in $\mathcal{P}$ and quasi-concave in $\rho_{x}$.  $\mathcal{U}$ is represented by $u_{n}$ and $u_{k}$ as in (3) and (4).

Firstly, we need to show the quasi-concavity and the continuity of the utility function of the PU. Furthermore, when $0<\mathbb{Q}_{N}<I_{n}+I_{k}$, the utility of the $nth$ PU is given as $u_{n}=\mathbb{Q}_{N}-\mu_{2}\left[\left(e^{\left(I_{n}+I_{k}-Q_{N}\right)}\right)\right]$, and the second order
derivative with respect to ${Q}_{N}$ is given by ${u}_{n}^{\prime \prime}=-\mu_{2} e^{\left(I_{n}+I_{k}-Q_{N}\right)})<0$, which is concave in $\mathbb{Q}_{N}$. On the contrary, when $\mathbb{Q}_{N}>I_{n}+I_{k}$, the utility of the $nth$ $\mathrm{PU}$ is given as ${u}_{n}=\mathbb{Q}_{N}-\mu_{1}\left[\left(\mathbb{Q}_{N}-I_{n}-I_{k}\right)^{2}\right]$, and the second order derivative of the $nth$ PU utility function is ${u}_{n}^{\prime \prime}=-2 \mu_{1}<0$ (i.e., ${u}_{n}$ is concave in $\mathbb{Q}_{N}$). Now, considering the case of $nth$ $\mathrm{PU},$ the best response of the utility function is a standard function. Here, the first derivative of this utility function is given by ${u}_{n}^{\prime}=1+$ $\mu_{2}\left[\left(e^{\left(I_{n}+I_{k}-Q_{N}\right)}\right)\right].$ Specifically, one gets negative utility when $\mathbb{Q}_{N}>I_{n}+I_{k}.$ 

On the other hand, when $\mathbb{Q}_{N}<I_{n}+I_{k_{n}}$, the first derivative of the utility function is given by ${u}_{n}^{\prime}=1-2 \mu_{1}\left[\mathbb{Q}_{N}-I_{n}-I_{k}\right],$ which has positive utility. As a result, the maximum utility of the $nth$ PU is achieved upon getting positive utility. Therefore, the best response of the $nth$ PU utility function is given by
\begin{equation}
    \mathbb{Q}_{n}^{*}=\frac{1}{2 \mu_{1}}+I_{n}+I_{k}, 
\end{equation}
note that: 
\begin{itemize}
  \item  $\mathbb{Q}_{n}^{*}>0$.
\item Given $\mathbb{P}_{1} \geq \mathbb{P}_{2},$ $\mathbb{Q}_{n}^{*}\left(\mathbb{P}_{1}\right)>\mathbb{Q}_{n}^{*}\left(\mathbb{P}_{2}\right)$. 
\item $ \forall \lambda, \lambda>1,$ then $\lambda \mathbb{Q}_{n}^{*}=\lambda \frac{1}{2 \mu_{1}}+\lambda I_{n}+\lambda I_{k}$ and
$\mathbb{Q}_{n}^{*}(\mathbb{P})=\frac{1}{2 \mu_{1}}+\lambda I_{n}+\lambda I_{k}.\text { Then, } \lambda \mathbb{Q}_{n}^{*}>\mathbb{Q}_{n}^{*}(\mathbb{P})$.
\end{itemize}

Consequentially, the best response of the PU utility function is a standard function.

Secondly, regarding to the SU, the power action sets of the SUs are closed subsets of $\mathbb{R},$ since the first condition is satisfied. Therefore, it is easy to verify that the utility functions of the SUs are continuous in $\rho_{k}.$ Hence, the best power solution of the $kth$ SU can be represented as follows
\begin{equation}
\rho_{k}^{*}=\frac{I_{k}\left(2^{\ln 2}-1\right)}{h_{s k}^{2}}, 
\end{equation}
where the concavity for the SUs can be improved as $\frac{d^{2} u_{k}}{d \rho_{k}^{2}}<0$. 
Based on that, the SU utility function is quasi concave. 

Algorithm 1 is used to solve the power control game. Here, $\mathbb{B}\left(\tau_{k}\right)$ denotes the set of the best transmit powers for the $\mathrm{SUs}$ at time instances $\tau_{k}$ in response to the interference vector $\rho_{-k}\left(\tau_{k}-1\right).$ Implementation of lower bound in the algorithm 1 can be done by assuming that the instantaneous SINR of the PU at the BS is known by the $kth$ SU. The $kth$ SU uses this information to derive its lower bound transmit power.
\begin{algorithm}
\begin{enumerate}
\item At time $t=0, Q(0)=Q$ for $P U$ at $n=1$. 
\par 
\item{For all terminals $\tau_{n} \in T$ .}{
\begin{itemize}
\item \textbf{Update} the power of the $nth$ PU using (11).\;
\item  \textbf{Compute}:$$\mathbb{B}\left(\tau_{n}\right)=\arg\\\max_{\mathcal{Q}_{n}\in\mathcal{Q}} u_{n}\left(\mathbb{Q}_{n}, \mathbb{Q}_{-N}\left(\tau_{n}-1\right)\right)$$
\item \textbf{Assign} $\mathcal{Q}\left(\tau_{n}\right)=\min \left(\mathbb{B}\left(\tau_{n}\right)\right).$ 

\item{At time $t=0,\rho(0)=\rho_1$ for the $\mathrm{SU}$ at $k=1$}.\;
\par
\item {For all $k$ such that $\tau_{k} \in T$.\;}
\item \textbf{Update} the power of the $kth$ SU using (12).\;
\item \textbf{Compute}:$$\mathbb{B}\left(\tau_{k}\right)=\arg\\\max_{\rho_{k} \in \rho} {u}_{k}\left(\rho_{k}, \rho_{-k}\left(\tau_{k}-1\right)\right)$$
\item \textbf{Assign} $\mathcal{P}\left(\tau_{k}\right)=\min \left(\mathbb{B}\left(\tau_{k}\right)\right).$ 
\end{itemize}
\item If $\mathbb{Q}_{N}<I_{n}+I_{k}$ then stop and don't accept any SU into the system model, else make $n=n+1$ then check for the SU.

Otherwise, if $\left\|\rho_{k}\left(\tau_{k}\right)- \rho_{-k}\left(\tau_{k}-1\right)\right\| \leq \epsilon,$ then stop and declare NE as $\rho\left(\tau_{k}\right)$ else, make $t=t+1$, return to step 2.
 }
\end{enumerate}
\caption{Best Response Power Control}
\end{algorithm}
 
\section{Simulation Results}
\begin{figure}[t]
\centering
\includegraphics[scale=0.5]{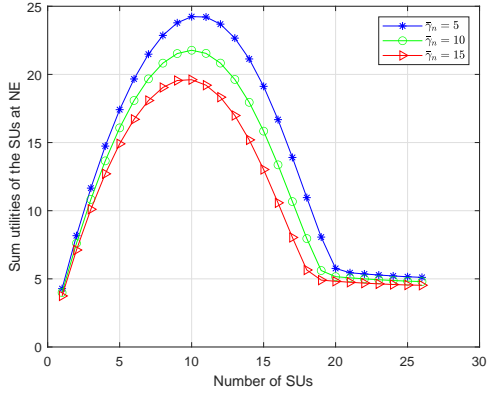}
\caption{The effect of minimum SINR of the PU on the summation of the utilities of SUs.}
\label{fig1}
\end{figure}
% {\color{red}{(as a minimum number of PU in any cognitive radio it will be 1 at least). Here, try to do the results to small number of SUs}}
In this section, we consider a CR-NOMA network based on different parameters which are represented in the Table I. The path attenuation between $kth$ SU and the secondary BS using the simple path loss model is $h_{ks}=\frac{0.097}{d_{ks}^{4}}$ \cite{peterson1995introduction}, where 0.097 approximates the shadowing effect.
\begin{table}[!h]
\caption {Simulation Parameters}
\label{title4}
\centering\renewcommand{\arraystretch}{1.40}
\resizebox{0.9\columnwidth}{!}{
\begin{tabular}{|c|c|}
\hline
{\textbf{Parameters}}                                                                         & {\textbf{Assumptions}} \\ \hline
{{Noise power}}                                                                        & 0.1   \\ \hline
{{Number of PUs}}                                                                      & 1    \\ \hline
{{Number of SUs}}                                                                      & 26    \\ \hline
{The target SINR for the nth PU}                                                              & 10                            \\ \hline
{The Pricing coefficients}                                                                    &   $\mu_{1}=10,\mu_{2}=100$                             \\ \hline
{The maximum transmission power for the SUs}                                                  & 20                            \\ \hline
{The bandwidth coefficient}                                                                   & 10                            \\ \hline
\multirow{4}{*}{{The distance}}                                                                 & [200,250,310,370,380,400, \\
          & 460,500,570,600,630,660,                  \\
  &  ,725,740,770,800,810,850,                  \\
       &  ,875,880,900,925,945,968,988,1000]                 \\ \hline
\end{tabular}}
\end{table}

% this is me okk I see ... I am shaima 
Fig. 1 shows the sum utilities of SUs versus number of SUs for different SINR thresholds, where each threshold shows unique maximum utility at $k=10$. When $k<10$, the decrease in the average SU utility is dominated by the increase of the number of SU in the system. However, the sum of the SU utilities still increases due to the increase in the number of SUs in the system. 
% Furthermore, the sum at the end doesn’t go to zero. The advantage of this utility can be represented as a steady-state summation of the utility.
In Fig. 2, the effect of different values of $\overline{\mathcal{P}}$ on the steady-state of these summations is analyzed. This steady-state summation is interesting for the users who are being served with the last resources in the system. When $\overline{\mathcal{P}}=13,$ it gives the maximum steady-state summation at the maximum number of SUs served in the energy-efficient mode. 
% $\overline{\mathcal{P}}=25,$ it gives the minimum steady-state summation as shown in Fig. 3. 
Fig. 3 and Fig. 4 offer the comparison of the NOMA and OMA \cite{li2008novel} approaches for SUs power at the NE. 
From Fig. 3, it should be noted that the OMA approach takes higher power at faster rate compared to NOMA. Higher power means a decrease in average utilities, and thus CR-NOMA can reduce the transmission power and the amount of interference in the network. 
\begin{figure}
\centering
\includegraphics[scale=0.5]{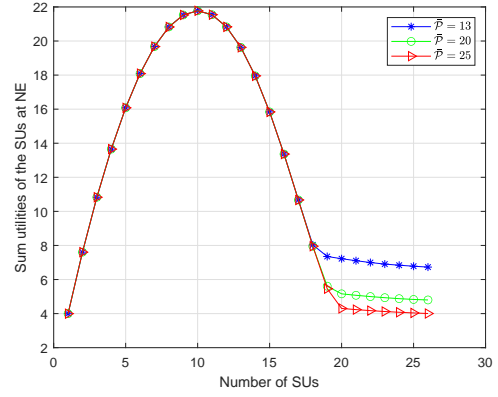}
\caption{The effect of maximum transmission power of the SUs on the summation of utilities.}
\label{fig2}
\end{figure}
\begin{figure}
\centering
\includegraphics[scale=0.5]{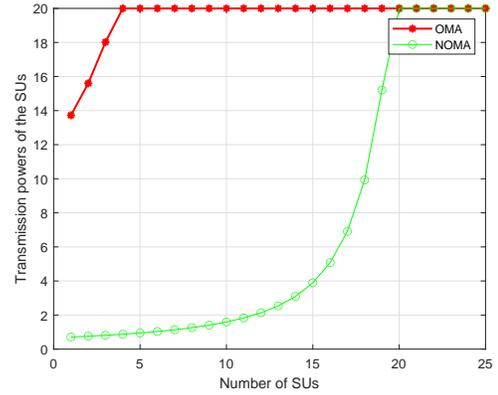}
\caption{The transmission power of SUs at the NE with NOMA and OMA.}
\label{fig3}
\end{figure}
\begin{figure}
\centering
\includegraphics[scale=0.5]{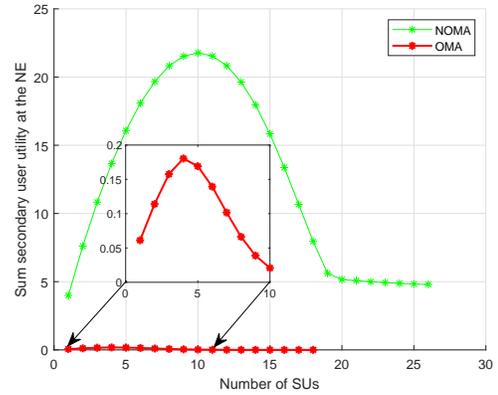}
\caption{Sum of the utilities of SUs at the NE with NOMA and OMA.}
\label{fig4}
\end{figure}
Fig. 4 shows the sum of the utilities of SUs at the NE with different approaches. Note that the OMA approach serves four users while the proposed NOMA accommodates at most $20$ users in an energy-efficient manner. Thus, the proposed approach gets more efficient results than OMA in terms of sum rate improvement and more user accommodation.

\section{Conclusion}

In this paper, the RA of CR-NOMA using a game-theoretic approach is studied. Several PUs coexist with the SUs using the game theory by modeling the natural interactions between the players in RA process. In particular, the power allocation problem is represented as a game, and NE is approached as the optimal power allocated to each user in the system. Finally, the superiority of the proposed scheme is shown through MATLAB simulation. The sum utilities of SUs with NOMA has significant improvements up to 37.5\% increase while compared with the OMA approach. Consequently, the maximum possible number of SUs in the energy-efficient mode that can be afforded in the system has increased up to 5.6\%. Based on that, Internet of Things (IoT) devices and Vehicle to Vehicle (V2V) communication could be a practical applications for the proposed system, where maximum capacity and efficient spectral sharing are met. In future works,  the EH for the cooperative system model will be considered.

\balance

%\bibliography{IEEEfull,shaima}
%\bibliographystyle{IEEEtran}


\end{document}